\documentstyle[12pt]{article}
\renewcommand{\baselinestretch}{2}
\begin{document}
\title{Why a Scalar Explanation of the L3
Events is Implausible \thanks{This work is supported
in part by funds provided by the U.S.
Department of Energy (DOE) under contract \#DE-AC02-76ER03069,
by the Texas National Research Laboratory Commission
under grant \#RGFY92C6 and by CICYT (Spain) under
Grant No. AEN90-0040.\hfill\break
\vskip -.25cm
\noindent $^{\dagger}$National
Science Foundation Young Investigator Award.\hfill\break
Alfred P.~Sloan
Foundation Research Fellowship.\hfill\break
Department of Energy Outstanding Junior
Investigator Award.\hfill\break
}}
\renewcommand{\baselinestretch}{1.0}
\author{Lisa Randall$^{\dagger}$
and Nuria Rius\\
Massachusetts Institute of Technology\\
Cambridge, MA 02139\\
}
\date{}
\maketitle
\renewcommand{\baselinestretch}{1.2}
\begin{center}{Submitted to: {\it Physics Letters} B}\end{center}
\vskip .2cm
\abstract{

We investigate the question of whether an
additional light neutral scalar can explain
the $l^+ l^- \gamma \gamma$ events
with high invariant mass photon pairs recently observed by the L3
collaboration.
We parameterize the low energy effects of the unknown dynamics
in terms of higher dimensional
effective operators.
We show that operators which allow for the scalar to
be produced and decay into photon pairs will
allow other
observable processes that
should have been seen in current experiments.

}
\vfill
March 1993 \hfill  MIT-CTP\#2191
\thispagestyle{empty}
\newpage

\newcommand{\beq}{\begin{equation}}
\newcommand{\eeq}{\end{equation}}
\def\bea{\begin{eqnarray}}
\def\eea{\end{eqnarray}}
\newcommand{\ba}{\begin{array}}
\newcommand{\ea}{\end{array}}

\section{Introduction}
The L3 collaboration has observed recently
an excess of $l^+ l^- \gamma \gamma$ events
with high invariant mass photon pairs, $M_{\gamma\gamma} \simeq 60$ GeV
\cite{l3}.
Both DELPHI and ALEPH have two similar events each, and OPAL seems
to have one candidate \cite{aspen}; however these experiments
are not so strongly peaked.
 Naively, such an observation would look like evidence of the discovery
of a new neutral particle. However, such an interpretation
requires a detailed understanding of the standard model background.
A recent calculation of the hard bremsstrahlung process
$e^+ e^- \rightarrow \mu^+ \mu^- \gamma \gamma$ \cite{kjo}
yields a significantly higher cross section at the $Z$ peak
than previous theoretical predictions had indicated, pointing
to the likelihood of a standard model explanation of the L3
events.  In this note, we address the question of the likelihood
of the discovery of a new scalar from a different vantage point;
we ask whether a nonstandard model with such a scalar
is consistent with other observational constraints.
We systematically investigate the possible scalar couplings
which could give rise to the L3 events
and show they are almost all excluded.

We  analyze the possibility of explaining the L3 events
\footnote {We will use only the L3 results (four events),
but our conclusions are not essentially changed
when we include the four LEP experiments.}
by assuming the existence of a light neutral scalar
($\phi$) of mass $m_\phi \simeq 60$ GeV. We assume the
scalar $\phi$ is a gauge singlet.
We parameterize low energy effects of unknown dynamics
at a scale $M$
in terms of higher dimensional effective operators,
constructed out of the Standard Model (SM) fields
and the extra neutral scalar.
These operators are suppressed by powers of
$M^{4-d}$, where $d$ is the dimension of the corresponding
operator. We impose
$SU(2)_L \times U(1)_Y$ gauge invariance, which
relates
the process of
interest with other observable effects which in general are
measurable and allows us to confirm or rule out the models.
We consider the most general possible low dimension operators
which can explain the observed events.  Our results do not
rely on assumptions on the expected size of the coefficients
of nonrenormalizable operators. We simply analyze other experimental
consequences of the operators which
could produce the observed events. In almost all cases this alone
is sufficient to rule out the operator.

In section 2 we briefly discuss
the possibility that
the new particle couples only to gauge bosons.
In section 3 we assume that
the scalar couples to leptons and gauge bosons
through lower dimensional gauge invariant
effective operators. In section 4 we analyze
higher dimensional four body
operators and in section 5 we summarize our
conclusions.

\section{Scalar Coupled to Gauge Bosons}

One can  consider a model in which the scalar $\phi$ couples
to the $Z$ boson and is produced
via the reaction
\beq
Z \rightarrow Z^* \phi  \ .
\label{p1}
\eeq

The case of $\phi$ being
the lightest CP-even neutral Higgs field ($h$)
in models with a non-minimal Higgs sector,
has been analyzed in ref. \cite{barger}.
To account for the fact that $h$ decays dominantly
to $2\gamma$, one can assume that $h$ does not couple to fermions,
but has essentially SM-type couplings to the $Z$ and $W$ bosons.
This can be achieved within the kind of models
referred to as model I in the literature \cite{gunion}, {\it i.e.},
models with two Higgs doublets of which only one couples to fermions.

However, this simplest possibility encounters immediately an
unavoidable problem which is independent of the particular model
for the scalar $\phi$.
The decays of the $Z$ boson are well known and the process
(\ref{p1}) would yield also final states
$\nu \bar\nu \gamma \gamma$,
with a branching ratio determined by the
well tested SM couplings of the $Z$ to neutrinos.
This implies that if we assume
$BR(Z \rightarrow Z^* \phi \rightarrow l^+ l^- \gamma \gamma)
= 4 \times 10^{-6}$ to account for the L3 events, we will
immediately obtain
$BR(Z \rightarrow Z^* \phi \rightarrow \nu \bar\nu \gamma \gamma)
= 8 \times 10^{-6}$.
The null results of searches for events with high
invariant mass photon pairs and missing energy
\cite{l3} translate into the upper limit
$BR(Z \rightarrow \nu \bar\nu \gamma \gamma) < 3 \times 10^{-6}$
at 95\% CL, if we assume (as we have done in all our estimates)
that the detection efficiency is one.
Therefore we conclude that the
$l^+ l^- \gamma \gamma$ events can not be explained by
the process (\ref{p1}).

\section{Lowest Dimensional Operators}

The next model we consider is one in which
$\phi$ couples to leptons, and
is therefore produced at LEP via
\beq
e^+ e^- \rightarrow Z \rightarrow l^+ l^- \phi \ .
\label{p2}
\eeq

The lowest dimensional gauge invariant operators involving the
scalar and two charged leptons are of dimension d=5, namely
\bea
O_{a} &=& {1 \over M} \bar{E_L} H e_R \phi \\
\label{oa}
O_{b} &=& {1 \over M} \bar{e_R} \gamma^\mu D_\mu e_R \phi \ ,
{1 \over M} \bar{E_L} \gamma^\mu D_\mu E_L \phi
\label{ob}
\eea
where $e_R$ refers to the right handed charged lepton,
$E_L$ is the $SU(2)_L$ doublet consisting of the charged
left handed lepton and the neutrino and $H$ is the standard Higgs
doublet.

We first consider $O_{a}$.
Notice that because
$\phi$ is an $SU(2)_L$ singlet, it has to include the
standard Higgs doublet.
When the standard Higgs gets a vev, $v$, this operator reduces to a
Yukawa coupling
\beq
{v \over M} \bar{e} e \phi \ .
\label{sll}
\eeq

We present two arguments against this model.
We first assume that the decay of $\phi$ into two photons is
induced by the gauge invariant effective  operator
\beq
O = {1 \over M'} \phi ( a B_{\mu\nu} B^{\mu\nu}
+ b  W^i_{\mu\nu}  W^{i \mu\nu} )
\label{sgg1}
\eeq
where $B_{\mu\nu}=\partial_\mu B_\nu - \partial_\nu B_\mu$,
$W^i_{\mu\nu}=\partial_\mu W^i_\nu - \partial_\nu W^i_\mu
+g f^{ijk} W^j_\mu W^k_\nu$
and $a,b$ are arbitrary coefficients.
When we write this operator in terms of the physical gauge fields,
there are three pieces involving the neutral gauge bosons
$Z$ and $\gamma$:
\bea
O_\gamma = {d \over M'} \phi F_{\mu\nu} F^{\mu\nu} & , &
d \equiv a c_w^2 + b s_w^2  \\
O_Z = {h \over M'} \phi Z_{\mu\nu} Z^{\mu\nu} & , &
h \equiv a s_w^2 + b c_w^2  \\
O_{\gamma Z}= {2k \over M'} \phi F_{\mu\nu} Z^{\mu\nu} & , &
k \equiv (b-a) c_w s_w
\label{sgg2}
\eea
where
$s_w$ ($c_w$) denotes the sine (cosine) of the electroweak mixing angle.


The two Feynman diagrams that contribute to
the process
$Z \rightarrow l^+ l^- \phi$
are depicted in Fig. 1. A tedious but
straightforward calculation leads to the following result for
the partial decay width:
\beq
\Gamma (Z \rightarrow l^+ l^- \phi) =
{\alpha \over s_w^2 c_w^2}
{1 \over 192\pi^2}
\left({v \over M} \right)^2
M_Z
[v^2 C_V(r) + a^2 C_A(r)]
\label{w1}
\eeq
where
$v$ ($a$) is the vector (axial) coupling of the lepton to the $Z$,
given by
\beq
v=-{1 \over 2} + 2 s_w^2 \ \ \
a=-{1 \over 2}
\eeq
and $r \equiv ({m_\phi \over M_Z})^2$.
The functions $C_V(r)$ and $C_A(r)$ take the form
\bea
C_V(r) &=& 2 r^2 F(r)-(1-2r-3r^2) \log r -2+8r-6r^2 \\
C_A(r) &=& -2 r^2 F(r)-(1+8r+3r^2) \log r \\
       &-&{11 \over 3}-5r+9r^2-{r^3 \over 3}
\eea
with
\beq
F(r)=2 L_2 \left({r \over 1+r}\right)
+ \log^2\left({r \over 1+r}\right)
-{1 \over 2} \log^2 r - {\pi^2 \over 6} \ ,
\eeq
and $L_2$ is the dilogarithm or Spence function.

L3 has a total of $\sim 10^6 Z$ events and 4
$l^+ l^- \gamma \gamma$ events have been observed, therefore
we assume
\beq
BR(Z \rightarrow l^+ l^- \phi) \times
BR(\phi \rightarrow \gamma \gamma) \sim 4 \times 10^{-6} \ ,
\label{exp}
\eeq
which yields a lower bound on ${v / M}$.
We obtain ${v/M} \ge 3.3$, and therefore $M \sim 75$ GeV
for $v=250$ GeV. This scale is so low that the validity
of treating this as an effective operator might be questioned.
This would be true particularly if one were to assume
the operator arose from a loop diagram in a more fundamental
theory.
Furthermore, this particle would have quite a large
width, on the order of its mass.  Presumably we should
not consider this operator further.  We nevertheless
show that such an operator
is decisively ruled out in any case
by current experimental data.

The experimental results on four fermion events at LEP
constrain the partial decay width of $\phi$ into two
photons to be at least of the same order of magnitude
as the corresponding decay width into leptons
\footnote{We thank B. Wyslouch for private communication.}.
In our model, these partial widths are, respectively:
\bea
\Gamma (\phi \rightarrow \gamma \gamma)&=&
\left({d \over M'} \right)^2
{m_\phi^3 \over 2\pi} \\
\label{wgg}
\Gamma (\phi \rightarrow l^+ l^-)&=&
3 \left({v \over M} \right)^2
{m_\phi \over 8\pi}
\label{wll}
\eea
where we have incorporated three lepton flavors.
We then impose $BR(\phi \rightarrow \gamma \gamma) \ge 0.5$.
This experimental constraint can be satisfied only if
$d^2 \sim 13$ when $M'=M$.  Notice that
taking $M'>M$ makes $d$ even
larger, well beyond the realm
of perturbation theory. We therefore
take $M'=M$ below.

Let us analyze now the remaining terms in eq. (\ref{sgg2}).
The operator $O_{\gamma Z}$ leads to the process
$Z \rightarrow \phi \gamma$ and the width is easily found to be
\beq
\Gamma ( Z \rightarrow \phi \gamma) =
{s_w^2 c_w^2 \over 6\pi}
{(b-a)^2 \over M^2}
\left({M_Z^2 - m_\phi^2 \over M_Z} \right)^3
\label{zsg1}
\eeq
Since $M$ can be no larger than determined
by eq. (\ref{exp}), the difference $|b-a|$ must be less than
$5 \cdot 10^{-2}$. This is determined since
$BR(Z \rightarrow 3\gamma)= BR(Z \rightarrow \phi \gamma)
\times BR(\phi \rightarrow \gamma \gamma)$, and
$BR(\phi \rightarrow \gamma \gamma) \ge 0.5$, so we obtain
$BR(Z \rightarrow 3\gamma) \ge 5 \times 10^{-2}(b-a)^2$.
The experimental upper limit for this process,
at 95\% confidence level, is
$BR(Z \rightarrow 3\gamma) < 1.4 \times 10^{-4}$ \cite{ggg}
and thus we conclude that
$|b-a| < 5 \times 10^{-2}$. This means that in order to be
consistent with the experimental data we have to assume
$b \sim a$ and therefore $d \sim h$ in eq. (\ref{sgg2}).

Finally, the last piece in eq. (\ref{sgg2}), $O_Z$,
contributes to the process $Z \rightarrow Z^* \phi$, which produces
final states of two fermions and the scalar when the $Z^*$ decays.
The expression for the branching ratio of the process
$Z \rightarrow \bar{f} f \phi$,
after the phase space integration,
is rather cumbersome, so we only give here the numerical result
for the neutrino decay channel:
\beq
BR (Z \rightarrow Z^* \phi
 \rightarrow \bar{\nu} \nu \phi) =
\left({h \over M} \right)^2 \cdot 3.5 \cdot 10^{-2} {\rm GeV}^2
\label{zzs1}
\eeq
Since $h \sim d$, this branching ratio is entirely determined.
Together with the constraint
$BR(\phi \rightarrow \gamma \gamma) \ge 0.5$ it implies that
$BR (Z \rightarrow \bar{\nu} \nu \gamma \gamma) \sim 4 \times 10^{-5}$,
which is excluded by LEP data \cite{l3}.

One can also rule out this model by considering the
rate for the scalar to be produced at TRISTAN.
The interaction (\ref{sll}) would also yield the direct
production $e^+ e^- \rightarrow \phi$.
The cross section for the process
$e^+ e^- \rightarrow \phi \rightarrow \gamma \gamma$
is easily found to be
\beq
\sigma(s) = {1 \over 4\pi}
\left({ d v \over M^2}\right)^2
{s^2 \over |s-m_\phi^2 + i m_\phi \Gamma_\phi|^2}
\label{t1}
\eeq
where $\Gamma_\phi$ is the total width of the scalar and
$\sqrt{s}$ is the center of mass energy.

Using that
$BR(\phi \rightarrow \gamma \gamma) \sim 0.5$ we obtain
$\sigma = 120$ nb,
which is inconsistent with current experimental data
($\sigma \sim 50$ nb for $\sqrt{s}=60$ GeV)
\cite{tristan}. This number was obtained within the
framework of this model, in which the scalar is  very broad.
The width determined at LEP would make the situation even worse.

We now consider the possibility
that the $Z$ decays into $\phi$ and two leptons directly
through the contact term $O_b$ in eq. (\ref{ob}).
Since the second of these operators
would yield
$Z \rightarrow \nu \bar{\nu} \phi$ with a branching ratio of the same
order of magnitude as
$Z \rightarrow l^+ l^- \phi$,
it is excluded.
We therefore assume this operator is
suppressed and restrict our attention to the
first one.

In terms of the physical gauge fields, $O_{b}$ is written as
\beq
O_b = {1 \over M} \phi \bar{e}_R \gamma^\mu
(\partial_\mu + ie A_\mu - ie {s_w \over c_w} Z_\mu) e_R   \ ,
\label{ob2}
\eeq
The last term
yields the following partial decay width for the process
$Z \rightarrow l^+ l^- \phi$:
\beq
\Gamma (Z \rightarrow l^+ l^- \phi) =
{\alpha \over 24 \pi^2}
{s_w^2 \over c_w^2}
{M_Z^3 \over M^2}
H(r)
\label{w2}
\eeq
where
\beq
H(r) = \left({3r^2 \over 16} + {r \over 4}\right)  \log r
+ {1 \over 8}
\left({3 \over 8} + {8r \over 3} - 3r^2 - {r^4 \over 24} \right)
\eeq
and $r=({m_\phi / M_Z})^2$.

Using again equation (\ref{exp}) derived from the L3 events, we
obtain
\beq
{1 \over  M^2} \sim 2.\times 10^{-3} \ .
\label{fb1}
\eeq
Although this result implies a very light mass scale ($M \sim 22$ GeV),
it depends also on the assumptions about
unknown coefficient in front of $O_b$. So we choose to study
the further consequences of the operator and we will show that it
can be excluded solely on an experimental basis.

The first term in $O_b$ yields also a derivative coupling $\phi l l$.
It is straightforward to compute the width for the decay
$\phi \rightarrow l^+ l^-$ induced by this coupling,
\beq
\Gamma(\phi \rightarrow l^+ l^-) = {m_\phi \over 16\pi}
\left({m_l \over M}\right)^2
\eeq
It is suppressed by the mass of the corresponding lepton, $m_l$,
and with the mass scale given by (\ref{fb1}) it turns out to be
$5.9 \times 10^{-10}$ GeV,
$2.6 \times 10^{-5}$ GeV and
$7.5 \times 10^{-3}$ GeV for $e$, $\mu$ and $\tau$ respectively.

The analysis done for the previous operators would not
apply here because of the helicity suppression in the
operator $O_b$.  In this case, the dominant $\phi$ decay
model is naturally to photons.  Furthermore, the
production cross section at TRISTAN would be quite small.
It is therefore best to rule out the operator directly,
by calculating the production cross section for three photons
with the mass scale $M$ we have already determined.

We look now at the piece of $O_b$
involving the photon field in eq. (\ref{ob2}).
This term of $O_b$ yields
$e^+ e^- \rightarrow \phi \gamma \rightarrow 3 \gamma$,
which can also be measured at LEP.
We obtain
\beq
\sigma(e^+ e^- \rightarrow \phi \gamma) = {\alpha \over 8 M^2}
\left({s-m_\phi^2 \over s}\right)
\label{g1}
\eeq
As the branching ratio of $\phi \rightarrow 2 \gamma$ is one in very
good approximation, we just have to plug in eq.(\ref{g1})
the value of $M$ determined by the L3 experiments to find
$\sigma(e^+ e^- \rightarrow 3 \gamma) \sim 0.4$ nb.
On the $Z$ pole, the peak cross-section for $Z$ production is
roughly 55 nb. Combined with the experimental upper limit
on the branching ratio for $Z \rightarrow 3 \gamma$ \cite{ggg},
we get
$\sigma(e^+ e^- \rightarrow 3 \gamma) < 8 \times 10^{-3}$ nb
and thus we conclude that the L3 events can not be due to $O_{b}$.

\section{Higher Dimension Four Body Operators}

If the neutral scalar is a singlet, the available
higher dimensional
gauge invariant operators have dimension d=7 and there are
three kinds of which we present three representatives
\bea
O_2 &=& {1 \over M^3} \bar{E}_L (D_\mu H) e_R D^\mu \phi  \\
\label{o2}
O_3 &=& {1 \over M^3} \bar{E}_L H (D_\mu e_R) D^\mu \phi  \\
\label{o3}
O_4 &=& {1 \over M^3} B_{\mu\nu}
\bar{e}_R \gamma^\mu e_R \partial^\nu \phi
\label{o4}
\eea
It is worth pointing out that these higher dimensional
operators do not contain any vertex involving only two fermions
and the scalar. Thus, in all these models $\phi$
decays dominantly into two photons, through the effective
operator $O$ introduced in section 3 (eq. (\ref{sgg1})).
Furthermore, unlike the lower dimensional operators
considered in section 3,
they can not be probed by direct production of
the scalar at TRISTAN.

Recall that in principle there are also operators analogous to
$O_4$ but involving the left handed $SU(2)_L$ doublets;
however those would once again yield the unobserved
$\nu \bar\nu \gamma \gamma$ events at LEP.

Notice that there can not be dimension 6 operators involving the Higgs
field because, as it is an $SU(2)_L$ doublet,
both $E_L$ and  $e_R$ are necessary
to make an $SU(2)_L \times U(1)_Y$ invariant
and by Lorentz invariance this implies that
two covariant derivatives are needed and therefore the lowest
dimensional operator has dimension d=7.

There are many other operators involving two covariant derivatives
as they can act on any couple
of the four fields involved, as well as both on the same field.
However, there is an essential difference
between $O_2$-type operators, in which one of the covariant
derivatives acts on the Higgs doublet, and operators of type $O_3$,
in which no covariant derivative acts on the Higgs.
The reason is that $O_2$-type operators only contain $Z$ and $W$
gauge bosons and we will show that
this fact prevents us from ruling them out with current
experimental data.
Operators of the $O_3$ kind involve also the photon and
thus they yield to $3 \gamma$ events at LEP with a cross section
directly dictated by the related $Z \rightarrow l^+ l^- \gamma \gamma$
branching ratio, as we have shown for the
lower dimensional operator $O_b$ in section 3.
The same applies to $O_4$.

In particular, the operator
$O_4$ defined
in eq. (\ref{o4}) has qualitatively
 the same consequences as the operator $O_b$.
In terms of the physical fields we have
\beq
O_4 = {1 \over M^3}
\bar{e}_R \gamma^\mu e_R \partial^\nu \phi
(c_w F_{\mu\nu} - s_w Z_{\mu\nu} )
\label{h3}
\eeq
The second piece leads to the process
$Z \rightarrow l^+ l^- \phi$ with a branching ratio
which agrees with the L3 results (eq. (\ref{exp})) for
$M \sim 74$ GeV. Then, we compute the cross section for
the process
$e^+ e^- \rightarrow \phi \gamma \rightarrow 3 \gamma$,
induced by the first term in eq. (\ref{h3}),
using the same mass scale. We obtain
$\sigma(e^+ e^- \rightarrow 3 \gamma) \sim 8 \cdot 10^{-2}$nb,
which is excluded by LEP data \cite{ggg}.

Similar conclusions will hold for other operators of this
sort. Of course there is a possibility of a fine tuned cancellation
among the many operators but this is even sillier than the model
already is.

Finally, we consider the operators of type $O_2$.
Since the results for all the operators of this kind are similar,
we present here only the detailed calculation for the operator in
eq. (\ref{o2}). When the Higgs field acquires a vacuum expectation
value, $O_2$ contains the piece
\beq
O_2^{nc} = {v \over M^3} {i e \over 2 c_w s_w} Z_\mu
\bar{e}_L e_R \partial^\mu \phi  \ ,
\label{nc}
\eeq
which leads to the following partial decay width for the process
$Z \rightarrow l^+ l^- \phi$:
\beq
\Gamma= {1 \over (4\pi)^3}
\left( {e \over 2 c_w s_w}\right)^2
\left( {v  \over M^3} \right)^2 M_Z^5 G(r)
\label{w3}
\eeq
where
\bea
G(r)& =& - {4 \over 15}
\left({1-r \over 2} \right)^5  \\
&+&
{1+r \over 6}
\left[
{(1-r)^3(1+r) \over 16}
-{3 r (1-r^2) \over 8}
-{3 r^2 \over 4} \log r \right]  \nonumber
\label{h1}
\eea
and $r \equiv {m_\phi^2 \over M_Z^2}$.
As in the previous models considered,
the branching ratio inferred from the L3 events
(eq. (\ref{exp})) provides an upper limit for the coupling,
\beq
{e \over 2 c_w s_w}
{v \over M^3} \sim 5.2 \cdot 10^{-4} \ ,
\label{h2}
\eeq
which implies $M \sim 56$ GeV.

The operator $O_2$ also induces $e^+ e^- \to \phi Z^*$,
which would produce photons and missing energy. However,
the rate is too small to be observable. The operator also
contains a charged current piece
\beq
O_2^{cc} = - {v \over M^3} {i e \over \sqrt{2} s_w} W_\mu
\bar{\nu}_L e_R \partial^\mu \phi
\label{cc}
\eeq
which induces the $W$ decay
$W^+ \rightarrow l^+ \nu_l \phi$. It is
straightforward to obtain that the width for this decay channel
is given by
\beq
\Gamma (W^+ \rightarrow l^+ \nu_l \phi) =
{1 \over (4\pi)^3}
\left( {e \over 2 s_w}\right)^2
\left( {v  \over M^3} \right)^2 M_W^5 G(r')
\label{w4}
\eeq
where
$r' \equiv {m_\phi^2 \over M_W^2}$ and the function $G$ was
defined in eq. (\ref{h1}).
When we incorporate in this expression the result (\ref{h2})
we get
$\Gamma(W^+ \rightarrow l^+ \nu_l \phi) \sim
2 \cdot 10^{-3}$ MeV,
which is not measurable in current and projected
experiments (it is expected that the total width of the $W$ boson
will be measured with a precision of 200 MeV at LEP II
\cite{ww}.).

We conclude that an operator of $O_2$ type cannot
be ruled out as decisively  as the others we have
considered.  It is however extremely unlikely that
it is responsible for the observed events.  First
of all, the scale of mass suppression is once again
too low to be really believable.  Furthermore, the
operator, if it existed, would most likely be chirally
suppressed. And finally, it would be
hard to understand why this operator should be induced
and not the others which we have successfully excluded.
We conclude that it is possible that a scalar
could be produced through this direct contact term
at the rate required, but it is extremely unlikely.

\section{Conclusions}

In this paper, we have considered the possibility that
there is a singlet scalar responsible for the observed two
photon invariant mass peak observed at LEP.  Of course,
there are more general possibilities one can consider.
For example, $\phi$ might have been part of an SU(2)
gauge multiplet. Presumably since the scale of the operators
is always very low, this will not matter since one can
insert the Higgs field (VEV) to make gauge invariant
operators and pursue an analysis identical to this one.
We  suspect methods similar to these will rule
out most particle models.

It might be objected that the scale of the operators
is always so low that we were not justified in
only considering the lowest dimension operators.
Again, with a more complete model of what induced
these operators one can mimic our analysis.
Given the full operator contributing to $\phi$ production and decay,
gauge invariance will
ensure that there are  related operators which lead to
three photon production at LEP or excess two photon
production at TRISTAN. Therefore, despite
the limitations of this approach, we anticipate
that the conclusion will be quite robust.

We conclude that it is very unlikely that the L3 events
represent the discovery of a new particle.  Even
without information on the angular distribution or
the standard model background,
we see that the events are not easily explained
in a particle physics model.

\begin{center}
{\bf Acknowledgements}  \\
\end{center}

We thank Bolek Wyslouch for motivating this investigation
and for useful discusssions.
This work is supported in part by funds provided by the U.S.
Department of Energy (D.O.E) under contract \#DE--AC02-76ERO3069
and by CICYT (Spain) under Grant No. AEN90-0040.
N.R. is indebted to the spanish MEC for a Fulbright scholarship.


\begin{thebibliography}{99}

\bibitem{l3}
L3 Collaboration, {\it Phys. Lett.} {\bf B295} (1992) 337.

\bibitem{aspen} J. Hilgart, talk given at the 1993
Aspen Winter Conference in Particle Physics, Aspen, Colorado (1993).

\bibitem{kjo}
K. Kolodziej, F. Jegerlehner and G.J. van Oldenborgh,
preprint BI-TP-93/01 (1993).

\bibitem{gunion}
J.F. Gunion, H.E. Haber, G.L. Kane and S. Dawson,
``The Higgs hunter's guide'' (Addison-Wesley, Redwood City, CA, 1990).

\bibitem{barger}
V. Barger, N.G. Deshpande, J.L. Hewett and T.G. Rizzo,
ANL-HEP-PR-92-102 (1992).


\bibitem{ggg}
DELPHI Collaboration,
{\it Phys. Lett.} {\bf B268} (1991) 296.


\bibitem{tristan} TOPAZ Collaboration
{\it Phys. Lett.} {\bf B284} (1992) 144.

\bibitem{ww}
P. Roudeau {\it et al.}, ECFA Workshop on LEP 200,
CERN report 87-08 (1987) 49.

\end{thebibliography}
\end{document}